\documentclass[aps,twocolumn]{revtex4}
\usepackage{graphicx}
\begin{document}

\title{\large \bf
Spin density wave and pseudogap in HTSC cuprates}
\author{\normalsize L. S. Mazov \\
\normalsize \it Institute for Physics of Microstructures,
\normalsize \it Russian Academy of Science\\
\normalsize \it Nizhny Novgorod, GSP-105, 603600, Russia }

\begin{abstract}
The evidence for the SDW nature of the pseudogap in HTSC cuprates
is presented from resistive and direct gap measurements (ARPES,
tunneling experiments etc.). Its d-wave symmetry mimics the
symmetry of the order parameter measured in the SC state. The
magnetic phase $H - T$ diagram for HTSC cuprates is formed. The
possible influence of an additional (SDW) order parameter on the
vortex structure in the applied magnetic field is analysed. The
behavior is discussed in terms of itinerant electron systems with
interplay between superconductivity and magnetism.
\end{abstract}
\maketitle
\section{\bf Introduction}

The nature of the pseudogap in the electron energy spectra of HTSC
cuprates observed in magnetic, neutron, optic etc. measurements
remains to be of interest during last decade. This pseudogap
appears in the normal state, well before the SC transition and
persists to liquid helium temperatures. Its $d_{x^2 - y^2}$-wave
symmetry seems to be attractive for possible d-wave
superconductivity (for review, see, e.g. \cite{Izyu1}). Though to
account for the formation of this pseudogap it was proposed a
number of models considered both the formation of Cooper pairs
above $T_c$ and the fluctuations of the antiferromagnetic (AF)
short-range order (for review, see, e.g. \cite{PS,OOP}) but the
nature of the pseudogap is an unresolved problem up to now.

\section{Experimental data and results of analysis}

\subsection{Evidence for SDW formation from resistive measurements
in the normal state}

To describe the normal-state behavior of HTSC cuprates there are now
proposed two main scattering mechanisms for mobile charge carriers: phonon
scattering (for review, see \cite{Maks}) and magnetic scattering ones
(see, e.g. \cite{Ito}), due to interaction of the carriers with AF spin
fluctuations.

In present work, it is assumed that total resistivity $\rho_{tot}$
in the normal state of HTSC cuprates includes both phonon
$\rho_{ph}$ and magnetic $\rho_m$ contributions. Such
decomposition ($\rho_{tot} = \rho_{ph} + \rho_m$) is usual for
magnetic metals (see, e.g.\cite{Vons}), and in \cite{M1} it was
proposed to use for HTSC cuprates. So determined magnetic
contribution $\rho_m$ in the whole temperature range of interest
may be extracted from the experimental $\rho(T)$ dependence by
subtracting the phonon term $\rho_{ph}(T)$.  But the extrapolation
of the linear part of $\rho(T)$ in paramagnetic (P) region (normal
state in HTSC) to $T = 0$ (that is usually performed for magnetic
metals), leads to substantial error in our case.  Because of this,
the general form of the Bloch-Gruneisen law should be used to
subtract $\rho_{ph}$
$$ \rho_{ph} = \rho_1 (\frac{T}{\Theta_D})^5
\int\limits_0^{\Theta_D/T} \frac{x^5}{(exp(x) - 1)(1 -
exp(-x))}dx,
\eqno(1)
$$
here $\rho_1$ is the scaling parameter.
The magnitude of $\rho_1$ may be estimated from the slope of the
normal-state region of the experimental $\rho(T)$ curve. It must
be noted that the slope of this linear region of the curve is here
considered as corresponding to the intermediate-temperature linear
region ($0.22 \le T/\Theta_D \le 0.43$), given by (1), which is
universal for conventional metals \cite{Zim}.
\begin{figure}
\includegraphics[width=8.0 cm]{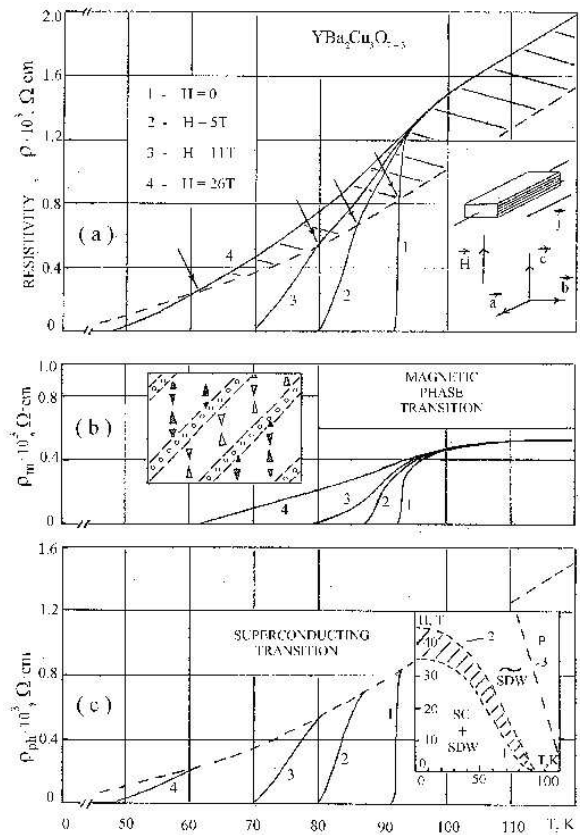}
\caption{Temperature dependence of in-plane resistivity in $YBCO$
(see, e.g. \cite{M1}) }\label{disp1}
\end{figure}

The $\rho_{ph}(T)$ curve corresponding to (1) with the parameters
of the $YBa_2Cu_3O_{7 - \delta}$ family of compounds is presented
in fig.1a (dashed curve). It is clearly seen from fig.1a that this
Bloch-Gruneisen curve intersects the experimental $\rho(T)$
dependences just at the "shoulder" (at $T = T_k$) and thus
separates the magnetic contribution $\rho_m$ to resistivity $\rho$
(shaded area in fig.1a) from the phonon one (see, also \cite{M1}
and references therein).

The temperature dependence of separated from each other magnetic
$\rho_m$ and phonon $\rho_{ph}$ contributions in the total
resistivity are presented in the fig.1b and 1c, respectively. From
fig.1b it is seen that there are two distinct regions in the
$T$-dependence of the magnetic contribution $\rho_m$. In the
normal state ($T > T_c(0)$) the magnitude of $\rho_m$ is high
enough and nearly independent of temperature. On the other hand,
at $T < T_c(0)$ the magnitude of $\rho_m$ gradually approaches
zero (sharp drop at $H = 0$), see fig.1b. This disappearance of
magnetic contribution $\rho_m$ at low temperatures is, in fact the
evidence for the appearance of some magnetic ordering in the
system, e.g. in the form of a modulated magnetic structure (cf.
with \cite{Izyu2}), say, the spin density wave (SDW) one formed in
the system near above $T_c$. So, two different magnetic phases
exist in the HTSC system at low and high temperatures,
respectively, and thus, a decrease of $\rho_m$ down to zero with
decreasing temperature may be considered as indirect evidence for
magnetic (SDW) phase transition.

It was proposed in \cite{M1} that such a magnetic (AF SDW)
ordering could be realized in the form of stripe structure in the
$CuO_2$-plane with periodical sequence of parallel charge and spin
(with AF arrangement) stripes alternating with each other (see,
insert in fig.1b). Namely, such a picture was obtained to that
time from both neutron experiments and numerical simulation in the
Habbard model (for more details, see \cite{Kato}). Nowadays fast
and local probes (ARPES, EXAFS, STM) (for details, see \cite{Bia})
demonstrate that this stripe structure arises well above $T_c$, in
the normal state (at $T \sim 150 K$, depending on the family of
compounds). Its arising is well correlated with the formation of
the pseudogap in the electron energy spectra of HTSC observed in
neutron, magnetic, optic etc. experiments (for review, see e.g.
\cite{PS,OOP}). This correlation between onset temperatures for
both pseudogap and stripes permits to propose a SDW nature of the
pseudogap in the HTSC cuprates.

Such a picture is supported by the magnetic phase $H-T$ diagram
for HTSC (see, insert in fig.1c) formed from the data of fig.1a.
The abscisses of "shoulder" points (fig.1a) being plotted  at the
$H-T$ plane fall in a straight line (open circles at the lower
part of dashed curve 2 in the insert of fig.1c) and being
considered as corresponding to the onset of the SC transition
($T_c^{onset}(H)$) form, in fact, temperature dependence of the
upper critical magnetic field $H_{c2}(T)$. On the other hand, the
dependence of $T_k(H)$ is, in fact, $T_m^{order}(H)$ curve
($\rho_m = 0$), see fig.1b. In this sense, curve 2 and curve 3
(which corresponds to the onset  of magnetic phase transition)
form the magnetic phase $H-T$ diagram for the given HTSC system.
So, below $T_k(H)$ curve the HTSC system appears to be in the
coexistence phase: SC and SDW, while above this $H_{c2}(T)$ curve
the system is in the nonsuperconducting phase of SDW  (modulated
magnetic structure whose wave number of modulation depends on both
$H$ and $T$ (cf. with \cite{Izyu2})). Above curve 3 the disordered
magnetic state is realized in the normal-state region of HTSC
cuprates.

\subsection{Evidence for SDW(CDW) origin of the pseudogap in cuprates
from direct gap measurements}

In the work \cite{Klemm} it was presented an evidence that the
HTSC pseudogap arises from CDW and/or SDW, and not from
thermodynamic fluctuations of the superconducting order parameter
phase. Such a conclusion was made on the basis of a detailed
analysis of angle-resolved photoemission spectroscopy (ARPES)
experiments, optical reflectivity, nuclear quadrupole resonance
(NQR), transport and thermodynamic measurements etc. It was noted
a striking similarity of the pseudogap regime in HTSC (with onset
temperature $T^*$) and in low-$T_c$ superconductors (LTSC), the
layered $2H-MX_2$ transition metal dichalcogenides, where M = Ta,
Nb and X = S, Se, below their respective incommensurate charge
density wave (CDW) transition temperaure $T_0$. The CDW in the
$2H-MX_2$ and HTSC pseudogap are considered as arising from
instabilities on the saddle bands (for more details, see
\cite{Klemm}) which are exhibited in the "normal" state of both
classes of materials. On the basis of comparison of ARPES
experiments for $Bi_2Sr_2CaCu_2O_{8 + \delta}$ (BSCCO) and for
$2H-TaSe_2$ (with highly anisotropic CDW) it was proposed that the
energy gap observed in ARPES experiments on BSCCO could have been
not pure a superconducting gap but that predominantly or combined
with the pseudogap  arising from CDW/SDW. It is noted the
complication of the search for the CDW/SDW appears due to defects
and due to the dynamical fluctuations of the CDW/SDW so that
inelastic neutron diffraction should be used to probe the dynamic
nature of the CDW/SDW. Then, since the charge density wave/spin
density wave is inherently dynamic in nature, it makes its direct
observation with electron diffraction ordinarily difficult as
well.

As for symmetry of the order parameter in HTSC, then this dynamic
gap formed by the CDW/SDW mimics the $d_{x^2 - y^2}$- wave
symmetry of the energy gap measured in the SC state. In this
connection, it was noted that many of superconducting properties
which had been ascribed to $d_{x^2 - y^2}$- wave superconductivity
may actually be due to the competing density wave (CDW/SDW) order
parameter which persists to low $T$.

Such a conclusion was then supported by recent Josephson junction
experiments with using of a novel, promising phase-sensitive
methods which measures namely SC order parameter characteristics
\cite{Li}. In that experiment, a very high quality single crystal
of BSCCO was cleaved in the $ab$-plane and then these two pieces
were rotated at a given angle about the $c$-axis and annealed. The
experiments with such a bicrystal $c$-axis twist Josephson
junction strongly evident to rule out a $d_{x^2 - y^2}$- wave
component in the superconducting order parameter. Moreover, it was
noted that the coexistence of a $d_{x^2 - y^2}$- wave and $s$-
wave order parameters (widely discussed in literature) is
incompatible with the crystal symmetry of BSCCO, since these order
parameters correspond to different representations of the crystal
groups.

\section{Discussion}

The above picture is in good agreement with the theory of
itinerant electron systems with interplay between
superconductivity and magnetism \cite{Mach}. In that theory, the
itinerant SDW gap may appear at the Fermi surface only before an
SC gap, i.e. in the normal state. This SDW gap is highly
anisotropic since it is only formed at symmetric parts of the
Fermi surface \cite{Mach} (see, fig.2).
\begin{figure}
\includegraphics[width=5.6 cm]{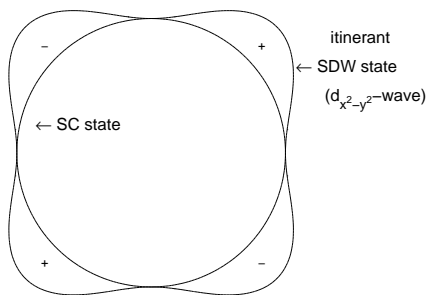}
\caption{Schematical sketch for coexistence of itinerant SDW and
SC states in HTSC}\label{disp2}
\end{figure}
Its width $\Delta_{SDW}$ being unusually large for the SC gap,
well conforms to that for the SDW gap because of inequality
$\Delta_{SC} < \Delta_{SDW}$ which is peculiar for the coexistence
phase in that model (see, fig.3). In such a case, the temperature
$T_c^{onset}$ (see, fig.1c) may be related to the appearance of
the SC gap which begins develop at the Fermi surface only when
transition of the HTSC system to a magnetically-ordered state is
over. Then, interrelations: $T_c^{onset} < T_{SDW}^{onset}$ (see,
fig.3) and $T_c^{onset} = T_m^{order}$ (see fig.1a) may be a
natural consequence of the equality of the magnetic ordering
energy $\varepsilon_m^{order}$ and the condensation pair one
$\varepsilon_c^{pair}$ considered as characteristic for such
itinerant electron system with an interplay between SC and
magnetism \cite{Mach}. Note here that namely such a picture was
recently reported from elastic neutron scattering experiments
\cite{Suz} when in La-based cuprate both the SC and static
magnetic order appear at the same temperature.

Then, the normal-state behavior of HTSC may be regarded in frames
of the theory of itinerant electron magnetism \cite{Moriya} (see,
also \cite{VT}). So, the independence of $T$ of a magnetic
contribution $\rho_m$ in resistivity may be described in terms of
so-called temperature-induced localized magnetic moments (LMM)
\cite{Moriya}. Such LMM, when both amplitude and orientation
fluctuate in time, arise in the itinerant electron systems because
of the saturation of amplitude fluctuations of local spin density
(FLSD) above some characteristic temperature $T^*$, where
transverse components begin prevail in FLSD spectra. On the other
hand, vanishing of magnetic contribution $\rho_m$ with decreasing
temperature may be considered (in frames of the above theory
\cite{Moriya}) as a consequence of the disappearance of transverse
components in the FLSD spectra of the HTSC magnetic subsystem. In
present treatment, this value of $T^*$ may be considered as the
onset temperature for the itinerant SDW ($T_{SDW}^{onset}$) state.

\begin{figure}
\includegraphics[width=5.6 cm]{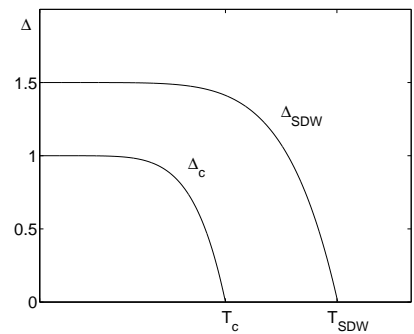}
\caption{Temperature dependence of SC and SDW order parameters in
HTSC cuprates (scheme)}\label{disp3}
\end{figure}

So, at $T > T^*$ ($\rho_m = const$, temperature-induced LMM
regime) the magnetoresistance effect is {\it negative} in sign (as
in paramagnetic state of usual magnets), and the influence of the
applied magnetic field is not strong (because of the linearity of
the Langevin formulae in this temperature region). On the
contrary, at $T < T^*$ (SDW regime), when the amplitude of FLSD is
out of saturation, the transverse magnetic field leads to a {\it
positive} magnetoresistance effect (increase in magnetic
contribution $\rho_m$). Such an effect is well known in AF
magnetic systems and is usually regarded as due to the enhancement
of spin fluctuations in the external magnetic field in such
systems (see \cite{M1}).

So, because of above, the characteristic temperature $T^* =
T_{SDW}^{onset}$, due to the temperature-induced LMM regime, is
not practically shifted by the applied magnetic field (cf. with
curve 3 in the insert in fig.1c) while the magnitude of the
itinerant SDW gap is essentially controlled by the intensity of
spin-fluctuation scattering depending on $H$ and $T$. Namely, such
a picture was observed in recent NMR measurements with YBCO
\cite{Gorny} where it was observed that the normal-state $(T >
T_c)$ "onset of pseudogap effects does not shift down in
temperature" up to $H = 14.8 T$ (though $T_c$ shifts down to 8K).
Such behavior in fact excludes SC fluctuations as origin of
pseudogap. The absence of any field effect is regarded in
\cite{Gorny} as indication to a relatively large energy scale
(characteristic for the itinerant SDW state rather than for the SC
one \cite{M2}, see also fig.3) for the pseudogap mechanism.

The magnetic field $H$ being applied along the $c$-axis
essentially determines the temperature behavior of the HTSC
system. In the normal state ($T > T_c^{onset}(H)$), because of the
enhancement of spin fluctuations (see above), it leads to a
broadening of resistive transition (due to the broadening of its
magnetic (SDW) phase transition part \cite{M1}). In the SC state
the dynamical stripe structure give rise to the intrinsic pinning
mechanism. Indeed, according to the general theory of SDW (for
review, see \cite{Izyu2,KT}), the formation of incommensurate SDW
in the system is accompanied by arising of CDW with wavelength
equal to $\lambda_{CDW} = \lambda_{SDW}/2$. In its turn, the
formation of CDW results in a wave of lattice deformation
(distortion) and thus in the formation of dislocation walls in the
transverse direction. Such walls provide, as known, an effective
pinning potential. Accordingly to above, the equivalent walls are
placed at the distance of near $5nm$, which value is determined by
$\lambda_{SDW} = 5nm$.  This value is also consistent with value
of $2\xi_{ab} = 5nm$ \cite{M2} which determines the vortex core
scale.

Such a picture is supported by STM measurements \cite{Hoog} where
it was observed that the vortex core has a complex spatial
structure and consists of two regions with a zero SC order
parameter. These regions are placed at the distance of a few tens
of angstroms, which fact is in well agreement with corresponding
distance between equivalent dislocation walls equal to
$\lambda_{SDW} = 5nm$.


\end{document}